\documentclass[slac_one]{revtex4}
\usepackage{graphicx}
\usepackage{fancyhdr}
\usepackage{cancel}

\begin{document}

\title{Discovery Potential for the SM Higgs Boson in the $H \rightarrow WW^{\ast} \rightarrow 2l2\nu$  channel at LHC}

\author{Rebeca Gonzalez Suarez (for the CMS Collaboration)}
\affiliation{U. de Cantabria, IFCA, Santander, Spain}

\begin{abstract}
A prospective analysis for the search of the Standard Model (SM)
Higgs boson with the CMS detector is presented in the context of
the early LHC data. The aim is to establish an analysis strategy
for inclusive production of the Higgs boson decaying in
$WW^{\ast}$ pairs in the context of the early LHC data. Higgs mass
region between 120-200 GeV, in which this signature was proposed
as highly sensitive ~\cite{examp2-ref} , has been studied. The $W$
decays into $l\nu$ are considered, where $l$ stands for $e$ or
$\mu$. The final states are characterized by two, opposite-sign,
high transverse momentum leptons, missing energy, carried out by
the undetected neutrinos, and little jet activity. This study uses
Monte Carlo (MC) events with full detector simulation, including
limited calibration and alignment precision as expected at the LHC
startup. Sets of sequential cuts are applied to each of the three
topologies, in order to isolate a signal which exceeds the
$t\overline{t}$ and continuum $WW$ backgrounds. Alternatively, an
artificial neural network multi-variate analysis technique is
used.
\end{abstract}

\maketitle

\thispagestyle{fancy}

\section{INTRODUCTION}

\subsection{Decay and production modes}

$H \rightarrow WW^{\ast}$ is the dominant Higgs decay mode in a
wide mass range, and for masses between $2m_{W}$ and $2m_{Z}$  the
Branching ratio is close to the unity.

SM Higgs in this mass range is mainly produced via two mechanisms,
Gluon fusion, in which case no hard jet activity is expected; and
Vector-boson fusion, characterized by 2 forward jets, opposite in
rapidity, with high mass.

The leptonic final states of the W bosons, electrons or muons,
give clear signature of the Higgs boson.

\subsection{Signal and background topology}

The signal, $H \rightarrow WW^{\ast} \rightarrow 2l2\nu$, presents
two leptons in the final state, with opposite charge and small
opening angle; and a significant missing transverse energy
($\cancel{E}_{T}$), due to the undetected neutrinos.

As no mass peak can be reconstructed, in this analysis the
background control is very important. The sources of possible
background are multi-lepton final states, specially final states
with $\cancel{E}_{T}$: di-boson production, especially $WW$,
$t\bar{t}$, $tW$, Drell-Yan and $W+jets$. When compared with the
signal, the cross-sections of the backgrounds are much larger.

\section{EVENT SELECTION}

\subsection{Trigger and lepton identification}

The amount of signal events recorded by the $CMS$ experiment will
depend on the trigger efficiency. A global $OR$ between different
HLT sequences (\textit{trigger-paths}) is chosen to maximize the
signal detection efficiency. The trigger-paths taken into
consideration require either a single lepton (electron or muon
with loose isolation requirements) or double leptons (electrons,
muons or a combination without isolation requirements) and they
have high efficiency at luminosities in the range $10^{31}$ to
$10^{33}$ $cm^{-2} s^{-1}$.


Standard $CMS$ lepton reconstruction techniques are used.

The effective selection of $e$ and $\mu$ show high efficiency for
true isolated leptons coming from $W$ boson decays, while at the
same time effectively suppressing leptons from heavy quark decays
or fake leptons produced by other objects. We use a tight
identification criteria for electrons due to the large $W+jets$
background, based on the matching of a charged track reconstructed
in the central tracker with a supercluster in the electromagnetic
calorimeter ~\cite{examp3-ref} ~\cite{examp4-ref}. Muon candidates
are identified by matching a track reconstructed in the muon
detectors with a track reconstructed in the central tracker
~\cite{examp4-ref} ~\cite{examp5-ref}.

\subsection{Lepton Selection}

Events are required to have exactly two leptons with opposite
electric charge sign, $|\eta|\leq 2.5$, $p_{T} \geq$ 10, 20 GeV,
and isolated (with tracks and in the calorimeter). If more than
two leptons fulfill the requirements, the event is rejected to
reduce $WZ$ and $ZZ$ contamination.

\subsection{Jet Veto and Missing $E_{T}$}

\begin{figure*}[t]
\centering
\includegraphics[width=60mm]{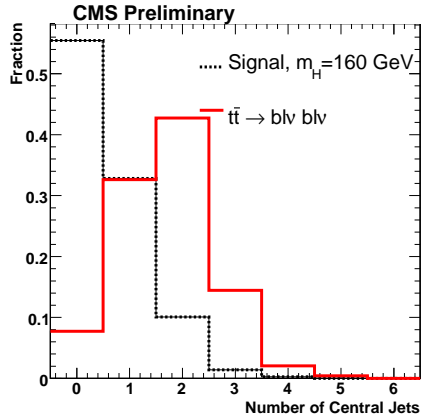}
\caption{Number of central jets for signal and $t\bar{t}$
background.} \label{JACpic2-f1}
\end{figure*}

No hard jet activity in the central region is expected for signal,
which can be used against $t\bar{t}$ background as shown in
Figure~\ref{JACpic2-f1}.

Jets are reconstructed using iterative cone algorithm with $\Delta
R$ = 0.5. If an event contains any jet with $p_{T} >$ 15 GeV and
$|\eta| <$ 2.5, it is rejected. The application of a jet veto is
also useful against: $tW$, $QCD$, $Z+jets$ and $W+jets$
backgrounds. Significant $\cancel{E}_{T}$ is expected due to the
neutrinos in the final state, and is computed from the raw
calorimeter tower energies, applying a correction due to the
presence of muons. The $\cancel{E}_{T}$ cut is specially useful
against Drell-Yan background after the invariant mass cut.

\subsection{Kinematic Variables}

To optimize the signal event selection against the main
backgrounds the following variables are used:

\begin{itemize}
    \item \textbf{angle between the leptons in the transverse plane
$\Delta \Phi_{ll}$}: for $WW$ events this angle is expected to be
large, for the scalar $SM$ Higgs boson this angle tends to be
small due to spin correlations.
    \item \textbf{invariant mass of the lepton pair $m_{ll}$}: an upper cut
is applied in the case of $e^{+}e^{-}$, $\mu^{+}\mu{-}$ final
states to reduce the contamination by leptons coming from Z-boson
decays.
    \item \textbf{transverse momenta of the harder ($p_{T}^{l,max}$) and
the softer ($p_{T}^{l,min}$) lepton}: upper/lower limit applied in
order to reduce the background further.
\end{itemize}

\begin{figure*}[t]
\centering
\includegraphics[width=70mm]{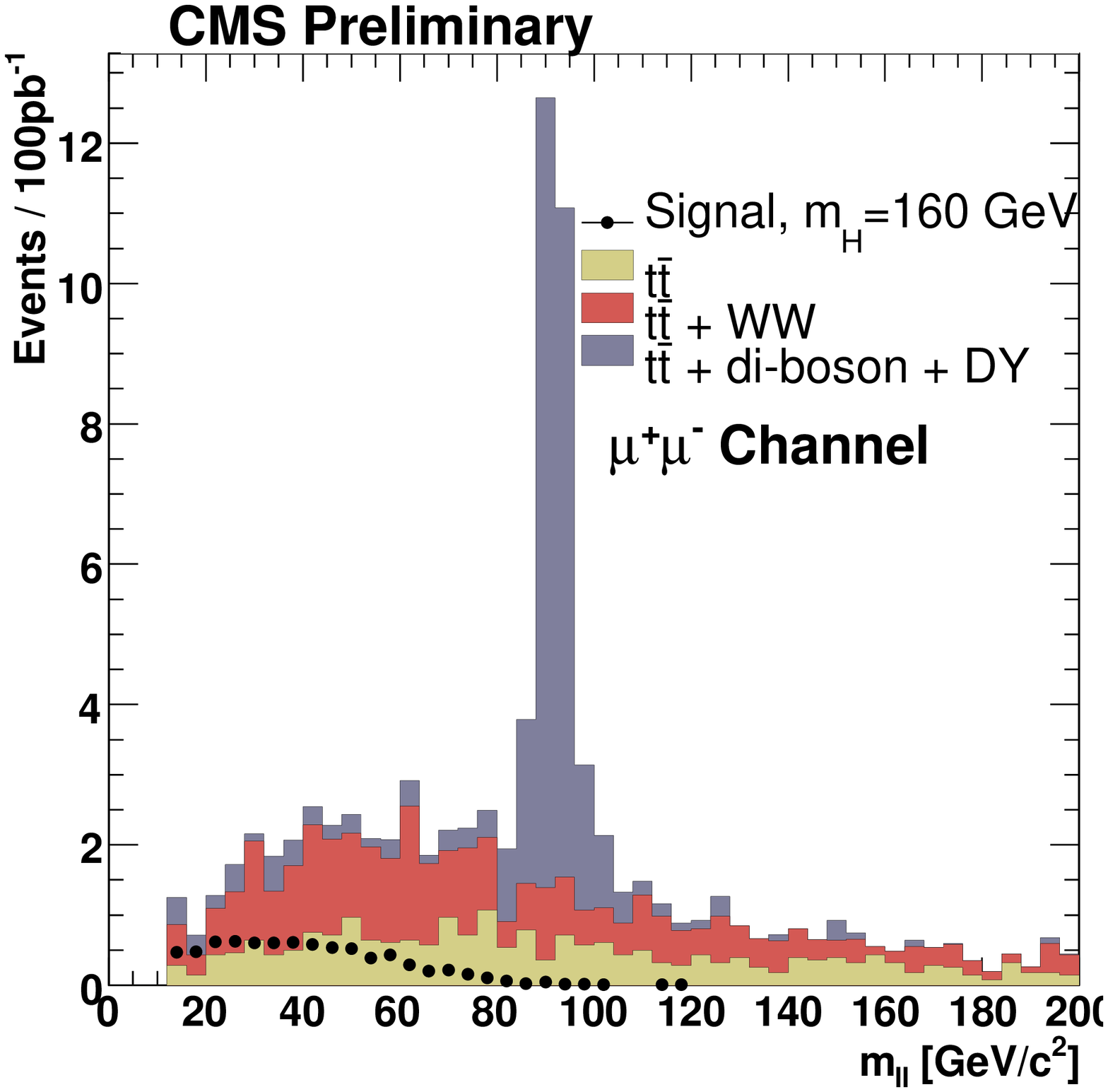}
\includegraphics[width=70mm]{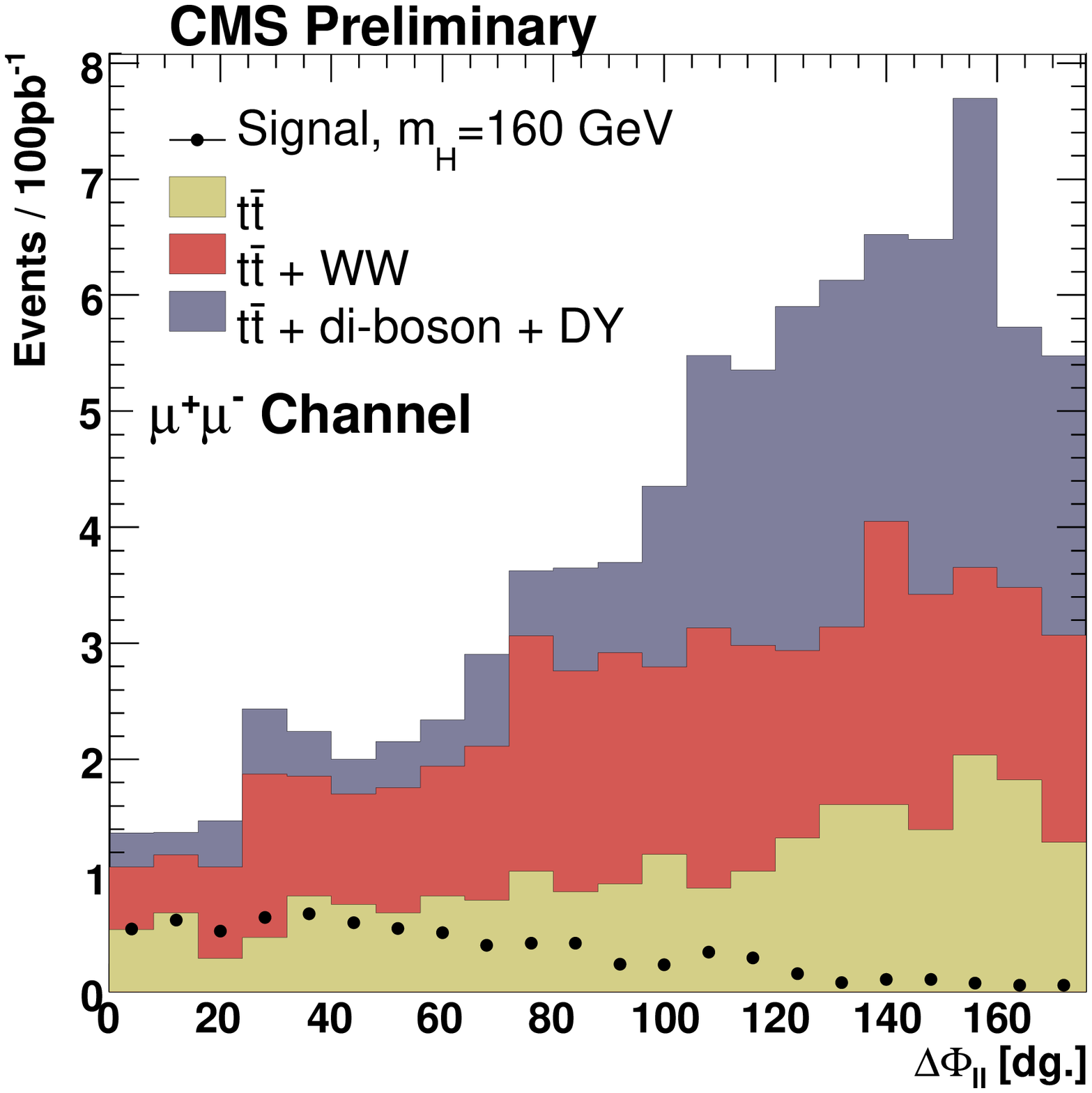}
\caption{Distribution of $m_{ll}$ (left) and $\Delta\Phi_{ll}$
(right) for signal and background in the $2\mu2\nu$ channel.}
\label{JACpic2-f2}
\end{figure*}

The distribution of the invariant mass $m_{ll}$ and the opening
angle $\Delta \phi_{ll}$ is shown in Figure~\ref{JACpic2-f2} for
the $m_{H}=160 GeV$ Higgs Boson signal and for the main background
in the $2\mu2\nu$ channel.

\section{ANALYSIS}

The analysis is split in three complementary topologies according
to the considered final states, $ee$, $e\mu$ and $\mu\mu$. A first
analysis based on sequential cuts was performed and then a
multivariate approach was also made.

The first part of the analysis is common for all the final states
and consist in the application of the \textbf{skimming} (to reduce
data and select potentially interesting events) and the
\textbf{pre-selection} (to define common objects of the analysis
-such as leptons or jets- and also to select events). Then, each
final state applies its own cuts to select events, due to the
experimental differences in the identification of $e$ and $\mu$
and also because some background processes do not equally affect
all the final states.

After the \textbf{Jet Veto}, the cuts on the sensible variables
(\textbf{$\cancel{E}_{T}$}, \textbf{$\Delta\Phi$},
\textbf{invariant mass} and the \textbf{$p_{T}$ of the leptons})
are applied. At the end, the results for the three final states
are combined to improve $CMS$ discovery potential.

\section{RESULTS}

The conclusion of the Physics $TDR$ ~\cite{exampl-ref} was that
the $SM$ Higgs boson could be discovered in the $H \rightarrow
WW^{\ast} \rightarrow 2l2\nu$ channel with less than $1fb^{-1}$ if
its mass is around 165 GeV. A new analysis taking fully into into
account the understanding of the detector and the control of
experimental and background systematics expected for intergrated
luminosities up to 1 fb$^{-1}$ has been  performed and progress
towards a complete and realistic event strategy is obtained


\begin{thebibliography}{9}   

\bibitem{exampl-ref} CMS Collaboration, "CMS Physics Technical
Design Report (V II)" J.Phys. G: Nucl.Part.Phys. 34,995-1579
(2007).

\bibitem{examp2-ref}
M.Dittmar and H.K.Dreiner Phys. Rev. D 55 (1997) 167.

\bibitem{examp3-ref}
S. Baffioni et al., "Electron reconstruction in CMS," CMS
NOTE-2006/040, and Eur.Phys.J. C49 (2007) 1099-1116.

\bibitem{examp4-ref}
CMS Collaboration, "The CMS experiment at the CERN LHC," S
Chatrchyan et al (2008) JINST 3 S08004

\bibitem{examp5-ref}
G. Bellan et al. Nucl. Phys. Proc. Suppl.177-178:253-254 (2008).

\end{thebibliography}
\end{document}